%% file: main.tex
\documentclass[aip,apl,reprint]{revtex4-1}
\usepackage[english]{babel}
\usepackage[utf8]{inputenc}
\usepackage{amsmath,amssymb,mathtools}
\usepackage{graphicx}
\usepackage[caption=false]{subfig}
\usepackage{siunitx}
\usepackage{enumitem}
\setlist{noitemsep}
\usepackage[hidelinks]{hyperref}

\graphicspath{{./figures/}}
\input{abbreviations}

\newcommand{\full}{$\mathrm{C_{60}}$}

\newcommand{\ie}{\textit{i.e.}}
\newcommand{\eg}{\textit{e.g.}}

\newcommand{\via}{\textit{via}}
\newcommand{\vs}{\textit{vs.}}

\begin{document}
\title{Estimating Evaporation Fields and Specific Heats Through Atom Probe Tomography}
\author{Andrew P. Proudian}
\author{Jeramy D. Zimmerman}
\email{jdzimmer@mines.edu}
\affiliation{Colorado School of Mines}
\date{\today}

\begin{abstract}
Estimations of evaporation field values in \gls{apt} literature are sparse despite their importance in the reconstruction and data analysis process. This work describes a straightforward method for estimating the \gls{zbef} that uses the measured voltage \vs\ laser pulse energy for a constant evaporation rate. This estimate depends on the sample radius of curvature and its \gls{cp}. If a similar measurement is made of the measured voltage \vs\ base temperature for a fixed evaporation rate, direct extraction of the material's \gls{cp} can be made, leaving only the sample radius of curvature as an input parameter. The method is applied to extract \gls{zbef} from a previously published voltage \vs\ laser pulse energy dataset for CdTe (\SI{18.07(87)}{\V\per\nm}); furthermore, using the published voltage \vs\ base-temperature sweep of CdTe permits extraction of a specific heat (\SI{11.27(254)}{\J\per\K\per\mol} at \SI{23.1}{\K}) in good agreement with the literature (\SI{11.14}{\J\per\K\per\mol} at \SI{22.17}{\K}). The method is then applied to the previously uncharacterized material \gls{irppy}, yielding $F_E = \SI{7.49(96)}{\V\per\nm}$ and $C_p = \SI{173(27)}{\J\per\K\per\mol}$; this \gls{zbef} is much lower than most materials characterized with \gls{apt} to date.
\end{abstract}

\maketitle
\glsresetall

As society demands ever more performant materials, detailed knowledge of nanostructure and composition becomes increasingly important. Unique among the many techniques for materials characterization, \gls{apt} combines high mass resolving power ($< \SI{0.5}{\dalton}$) and high spatial resolution ($< \SI{1}{\nm}$) in the same analysis.%
\cite{Amouyal2016}

In \gls{apt}, a sample is prepared with a small radius of curvature and a high bias is applied; this leads to a field at the surface of the sample of $>\SI{1}{\volt\per\nm}$. By applying an appropriate voltage or laser pulse, a single atom or molecule at the surface can be field ionized. The pulse-to-detection time and knowledge of the electric field allows the ion's mass-to-charge ratio and surface position to be determined. Repeating this process permits a three-dimensional, mass-specific reconstruction to be created.

When evaluating materials and understanding the evolution of an \gls{apt} sample, it is important to know the evaporation field(s) of the surface atoms or molecules. This is useful both for the practical reason of improving the quality of reconstructions (\eg\ through simulation), but also more broadly to clarify the physical process of field evaporation.%
\cite{Oberdorfer2013,Vurpillot2016}
Typical evaporation field values for elemental materials range from \SIrange{20}{150}{\volt\per\nm},%
\cite{Muller1969,Gault2012}
but there are few tabulated values for more complex materials, with most being of solute atoms in a metallic matrix.%
\cite{Brandon1966,Muller1969}
As the \gls{apt} community continues to expand the classes of materials it studies---such as ceramics, compound semiconductors, and organic small-molecules---the evaporation properties of these materials must be understood.%
\cite{Joester2012,Amouyal2016,Proudian2016,Proudian2019}

Due to the complexity of these new (for \gls{apt}) material systems, a framework for experimentally determining the evaporation fields of materials is necessary, as computational methods struggle to work in these high-field regimes (to say nothing of complex molecular systems).%
\cite{Larson2013}
Furthermore, the experimental method must be fast and easy to perform so that it is adopted and evaporation field values become more regularly reported.

In this work, we propose a method to estimate the evaporation field based on a simple series of voltage \vs\ laser pulse energy measurements at a constant evaporation rate. These provide the necessary constraints on the governing equations to extract the \gls{zbef}. We find that in a sample of CdTe for which data has been previously published our method extracts
$F_E = \SI{18.07(87)}{\volt\per\nm}$;%
\cite{Diercks2015}
this agrees with \gls{zbef} found on these data previously using the self-consistent reconstruction framework.%
\cite{Diercks2018}
In addition, a measurement series of voltage \vs\ base temperature at a constant evaporation rate%
\cite{Diercks2015}
recovers the \gls{cp} of the material ($C_p = \SI{11.27(254)}{\J\per\K\per\mol}$) in good agreement with direct measurements.%
\cite{Birch1975}

This method enables new measurements of \gls{zbef} and \gls{cp} for the organic small-molecule \gls{irppy}, which has $F_E = \SI{7.49(96)}{\V\per\nm}$ and $C_p = \SI{173(27)}{\J\per\K\per\mol}$.

To arrive at these estimates of \gls{zbef}, we begin with the basic assumption that the field strength at the surface is given by%
\cite{Larson2013}
$$F=\frac{V}{R k_f},$$
where where all variable definitions are given in \autoref{tab:vars}. If the ionization process is thermally activated, the rate of ionization is%
\cite{Larson2013}
$$r=A n_{\text{hr}}
\exp \left(-\frac{Q_n(F)}{k_B T}\right).$$

To estimate \gls{zbef}, we assume ionization is a purely thermal process%
\cite{Kellogg1981}
and that the function of the laser is to heat the tip. This assumptions allows us to change the denominator of the exponential as $k_B T \to k_B T +\alpha\delta E_p$. For the numerator of the exponential, close to \gls{zbef} we can assume the field sensitivity is linear, taking the form
$Q_n = \beta_n(F-F_E)$ because of the definition that $Q_n(F_E)\equiv0$.%
\cite{Marquis2008,Larson2013}
With these simplifying assumptions, the rate equation becomes
\[r=A n_{\text{hr}} \exp \left(-\beta_n\frac{\frac{V}{k_f R}-F_E}{k_B T+\alpha \delta E_p}\right).\]

Assuming a gaussian beam with a circular cross-section, this means that
\[\alpha (R)=1-\exp \left(-\frac{1}{2} \left(\frac{R}{\sigma }\right)^2\right).\]

With a sample detector distance of $L$ and a detector radius for collected ions of $r_d$, we estimate the field of view assuming point projection from a spherical tip. The maximum angle is $\theta = \tan^{-1}\frac{r_d}{L}$, meaning that the visible area ($A_s$) is
\[A_s = 2 \pi R^2 \frac{r_d/L}{\sqrt{1+(r_d/L)^2}}.\]
Approximating an atom or molecule as having a circular cross-sectional area with radius $d$ results in
\[n_{\text{hr}}=\eta\xi\frac{2(r_d/L)}{\sqrt{(r_d/L)^2+1}}\left(\frac{R}{d}\right)^2,\]
where $\eta$ is the detection efficiency and $\xi$ is the fraction of high probability surface molecules.

%

These simplifications result in the rate equation we will use to estimate \gls{zbef}:
\begin{equation}
\begin{split}
r =& A \eta\xi\frac{2(r_d/L)}{\sqrt{(r_d/L)^2+1}}~\times \\
&\left(\frac{R}{d}\right)^2
\exp \left(-\beta_n \frac{\frac{V}{k_f R}-F_E}{k_B T+\alpha \delta E_p}\right).
\end{split}
\label{eq:rate}
\end{equation}

Using \autoref{eq:rate}, we know that for measurements on the same sample
\[\frac{V-V_E}{k_B T + \alpha\delta E_p} = const.,\]
where we have converted $V_E=F_E k_f R$ because these are measured on the same sample and we assume the radius does not change over the course of the measurements (or can be corrected).%
\cite{Diercks2015}
This means we can write
\begin{equation}
V = C\alpha\delta E_p + C k_B T + V_E.
\label{eq:v-fit}
\end{equation}
Thus, a fit to the form $ax+b$ on a measurement series of voltage \vs\ pulse energy allows us to extract $C=\frac{a}{\alpha\delta}$ and therefore
\[V_E = b - \frac{a}{\alpha\delta}k_B T.\]

Given the definition of $\delta$ in \autoref{eq:rate}, \gls{cp} of the material is given by
\[\delta = \frac{k_B}{C_p N_m},\]
where $N_m$ is the number of moles of the material.
Assuming both a spherical interaction volume of the laser with the sample that matches the sample radius of curvature $R$ and a spherical atom or molecule with radius $d$, the measurements can be used to calculate \gls{cp} for the material under study:
\begin{equation}
\delta = \frac{k_B N_A d^3}{C_p R^3}.
\label{eq:heat-capacity}
\end{equation}
This means that \gls{zbef} becomes:
\begin{equation}
F_E = \frac{1}{k_f R}\left(b-\frac{a C_p R^3 T}{\alpha N_A d^3}\right).
\label{eq:f-fit}
\end{equation}

The best way to estimate $\delta$ in \autoref{eq:v-fit} is by measuring $V(T)$; that is, measure how voltage changes with temperature while holding the evaporation rate and pulse energy constant. We can calculate a value of $\delta$ by setting the two values of $V_E$ equal, which is allowed because of the assumption of constant $R$:
\[\delta =\frac{b_T - b_E - \sqrt{\left(b_T-b_E\right)^2 + 4 a_E a_T E_0 T_0}}{2 a_T \alpha E_0 / k_B},\]
where the subscripts denote whether the coefficient is derived from the energy ($E$) or temperature ($T$) fit. Using \autoref{eq:heat-capacity} this can be used to measure \gls{cp} for a material.

\begin{table}
\begin{tabular}{cp{2.7in}} 
\hline\hline
Variable & Description \\ \hline
$F_E$ & zero-barrier evaporation field \\
$V$ & sample voltage \\
$R$ & tip radius \\
$k_f$ & field factor \\
$k_B$ & Boltzmann constant \\
$T$ & sample temperature \\
$\alpha$ & sample cross-section fraction \\
$Q$ & field sensitivity \\
$\beta$ & field sensitivity linear coefficient \\
$\delta$ & pulse energy conversion factor \\
$L$ & sample-detector distance \\
$r_d$ & detector radius \\
$A_s$ & detector visible area \\
$d$ & atomic/molecular radius \\
$\eta$ & fraction of surface atoms/molecules likely to evaporate \\
$\xi$ & detection efficiency \\
$r$ & evaporation rate \\
$E_p$ & laser pulse energy \\
$C_p$ & specific heat \\
$N_A$ & Avogadro's number \\
$N_m$ & number of moles of the material \\
$a_E$ & slope of the linear fit on pulse energy \\
$a_T$ & slope of the linear fit on base temperature \\
$b_E$ & intercept of the linear fit on pulse energy \\
$b_T$ & intercept of the linear fit on base temperature \\
\hline\hline
\end{tabular}
\caption{The variables used for this development.}
\label{tab:vars}
\end{table}

\begin{figure}
\subfloat{\includegraphics[width=\columnwidth]{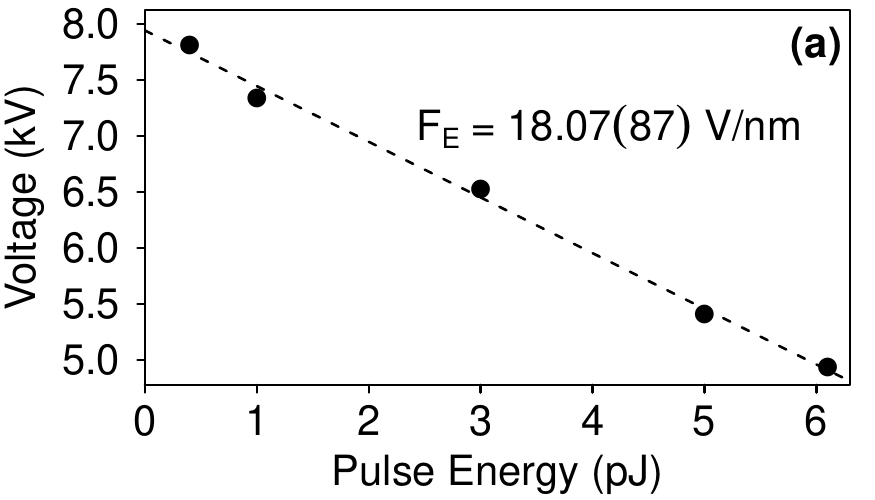}} \\
\subfloat{\includegraphics[width=\columnwidth]{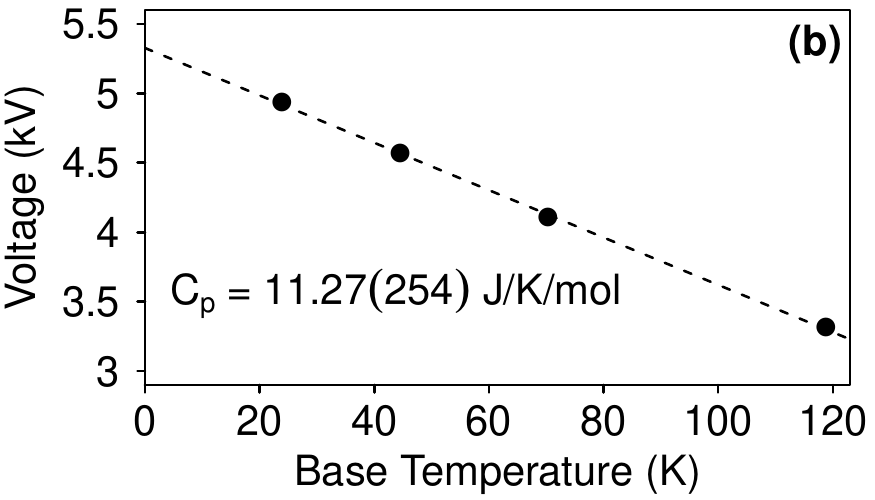}}
\caption{
\textit{(a)} The univariate fit of voltage \vs\ pulse energy for the CdTe data in \citeauthor{Diercks2015}.%
\cite{Diercks2015}
\textit{(b)} The univariate fit of voltage \vs\ base temperature for the CdTe data.
}
\label{fig:cdte}
\end{figure}

We test our method using CdTe, a well-characterized material.
In \citeyear{Diercks2015}, \citeauthor{Diercks2015} measured voltage \vs\ pulse energy and voltage \vs\ base temperature curves for CdTe, showing linear relationships in both series (see \autoref{fig:cdte}) that are consistent with the above development.%
\cite{Diercks2015}
For their data, the static radius assumption is fulfilled by the corrected voltage used to generate the curves, in which they accounted for the changing radius by correcting based on measured flux and \gls{tem} measurements of the initial and final radius of the sample.
This means that they set a reference tip shape and corrected subsequent datasets back to this shape; for the CdTe data, this is a radius of \SI{140}{\nm}.

Using the literature value of $C_p = \SI{11.14}{\J\per\K\per\mol}$ for CdTe at \SI{22.17}{\K},%
\cite{Birch1975}
the fit gives $F_E = \SI{18.07(87)}{\V\per\nm}$.
This agrees with \gls{zbef} previously reported for CdTe ($F_E = 12.4 ^{+7.4}_{-4.3}\mathrm{V~nm^{-1}}$), but with significantly smaller uncertainty because we do not rely on the self-consistent reconstruction process and instead fit directly on the voltage \vs\ pulse energy series.%
\cite{Diercks2018}
Fitting the voltage \vs\ base temperature curve using a value for $d$ in \autoref{eq:heat-capacity} of the distance between Cd and Te in a zinc blende crystal with a lattice constant of $a = \SI{0.648}{\nm}$ gives $C_p = \SI{11.27(254)}{\J\per\K\per\mol}$ for CdTe at \SI{23.1}{\kelvin}, which compares quite favorably with the literature value.%
\cite{Birch1975}
These two comparisons suggest that the assumptions of our development are reasonable and provide accurate estimates of \gls{zbef} and \gls{cp}.

With these results obtained on CdTe matching the literature, it confirms our technique and permits us to move on to a new material. 
We fabricated a sample with a \SI{102.4}{\nm} thick film (measured by spectroscopic ellipsometry on a silicon witness) of \gls{irppy} onto a silicon tip having a radius of $R=\SI{256}{\nm}$ following the procedure outlined in \citeauthor{Proudian2019}.%
\cite{Proudian2019}
Each voltage measurement is the average voltage for evaporation of \SI{5e4}{ions} at a fixed detection rate of \SI{0.3}{\percent}.
The first \SI{1.9e6}{ions} were omitted for sample alignment and radius equilibration.

\begin{figure}
\centering
\includegraphics[width=\columnwidth]{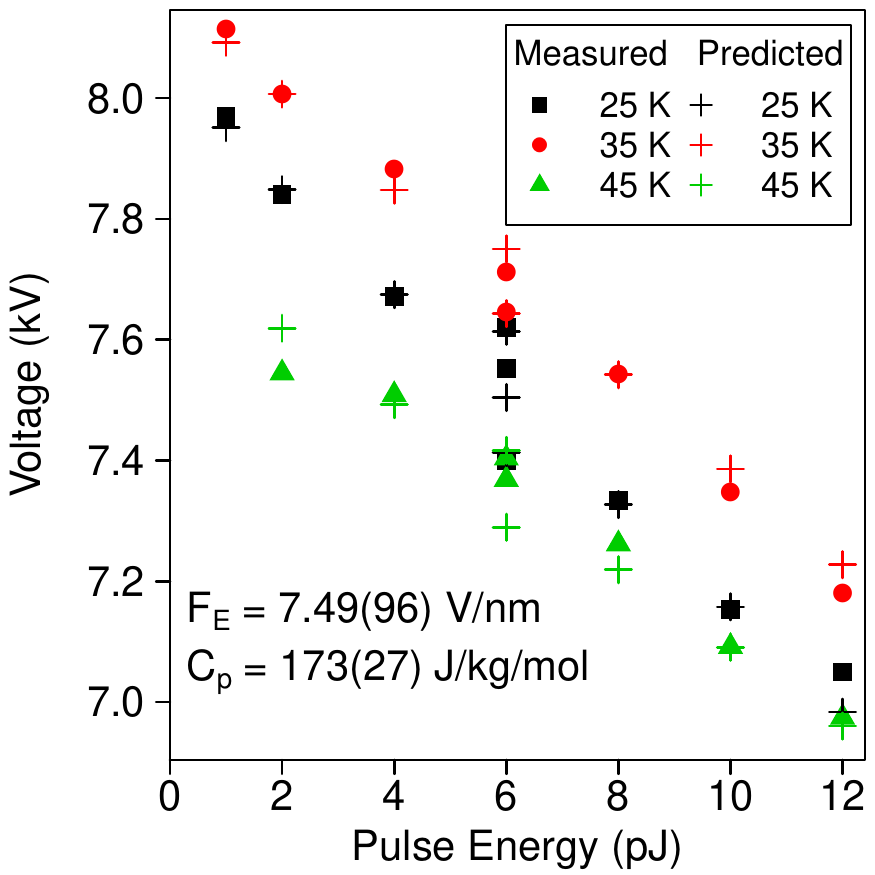}
\caption{Equilibrium evaporation voltage of \gls{irppy} \vs\ laser pulse energy, with different base temperatures shown in color. A multivariate fit following \autoref{eq:v-fit} using the radius assumption in \autoref{eq:radius} gives $F_E = \SI{7.49(96)}{\V\per\nm}$ and $C_p = \SI{173(27)}{\J\per\K\per\mol}$, with other fitting parameters given in \autoref{tab:irppy}; the predicted voltage values based on this fit are shown.}
\label{fig:irppy}
\end{figure}

The base temperature was set to \SIlist{25;45;35}{\K} and then pulse energies were sequentially adjusted as \SIlist{6;12;10;8;6;4;2;1;6}{\pico\joule} to generate the data shown in \autoref{fig:irppy}. \SI{4.9e6}{ions} were collected from the sample in total.

Fitting of this multivariate data to \autoref{eq:f-fit} was performed using non-linear least-squares fitting \via\ the \texttt{nls} function in the R package \texttt{stats}. The radius change of the sample is assumed to have the form
\begin{equation}
R = r_0 + r_f (1 - \exp(-N/n_r)),
\label{eq:radius}
\end{equation}
where $r_0 = \SI{256}{\nano\meter}$ is the initial sample radius, $r_f$ is the final sample radius increase, $N$ is the number of evaporated ions, and $n_r$ is an ion-radius constant. With this adjustment, the predicted voltage values based on this fit are shown in \autoref{fig:irppy}; the values of the fitting parameters are given in \autoref{tab:irppy}.

\begin{table}
\begin{tabular}{cl}
\hline\hline
Parameter & Value \\ \hline
$F_E$ & \SI{7.49(96)}{\V\per\nm} \\
$C_p$ & \SI{173(27)}{\J\per\K\per\mol} \\
$r_f$ & \SI{114(49)}{\nm} \\
$n_r$ & \SI{1.95(57)e6}{Ion} \\
$C$   & \SI{1.510(94)e12}{\V\per\J} \\
\hline\hline
\end{tabular}
\caption{The fitted values for the \gls{irppy} data shown in \autoref{fig:irppy} based on a multivariate fit following \autoref{eq:v-fit} using the radius assumption in \autoref{eq:radius}.}
\label{tab:irppy}
\end{table}

The linear behavior of both the CdTe and \gls{irppy} samples suggests our thermal ionization assumption is valid for these materials. In contrast, the non-linear behavior of GaN observed by \citeauthor{Diercks2015} suggests a different evaporation mechanism and hence is not amenable to this method of evaporation field extraction.%
\cite{Diercks2015}

Applying \gls{apt} to organic small-molecule semiconducting materials is just beginning, and there is much to learn about how these materials behave during field evaporation.%
\cite{Joester2012,Proudian2016,Proudian2019}
The measured evaporation field of \gls{irppy} is only $F_E = \SI{7.49(96)}{\V\per\nm}$, which is significantly lower than the fields typically measured for inorganic materials even though the mass of \gls{irppy} (\SI{654.78}{\dalton}) is much higher than the ions commonly observed in \gls{apt} samples.%
\cite{Muller1969,Gault2012}
This can be understood by considering that these solids are bound together by van der Waals bonds that are much weaker than the metallic or covalent bonds of the materials that are more commonly analyzed with \gls{apt}.%
\cite{Chickos2003}
These weak van der Waals bonds have correspondingly low evaporation fields, which is why entire molecules in this class of materials can field evaporate without fragmenting when analyzed with \gls{apt}.%
\cite{Proudian2019}

The inversion of the equilibrium voltage curves for the data at \SIlist{25;35}{\K} can be understood by recognizing the order in which these data were acquired and the consequent change in the sample radius of curvature ($R \approx$ \SIrange{340}{348}{\nm} at \SI{25}{\K} and $R \approx$ \SIrange{357}{361}{\nm} at \SI{35}{\K}); we accounted for this change using \autoref{eq:radius} when estimating \gls{zbef} and \gls{cp}.

Comparing the measured specific heat of $C_p = \SI{173(27)}{\J\per\K\per\mol}$ for \gls{irppy} with the measured value for \full\ at these temperatures of $C_p = \SI{50}{\J\per\K\per\mol}$ shows a comparable scale of \gls{cp} but higher, which is in line with the increased internal degrees of freedom for \gls{irppy}.%
\cite{Matsuo1992}
This \gls{cp} appears constant over the range of temperatures measured (\SIrange{25}{45}{\K}); \ie, adding in a linear slope correction to the fitting function results in a poorer quality model as tested using Schwarz's Bayesian Criterion.%
\cite{Sakamoto1986}
This behavior is similar to the relatively flat region of \gls{cp} for \full\ observed by \citeauthor{Matsuo1992} in this temperature range.%
\cite{Matsuo1992}

In conclusion, through \gls{apt} measurements of voltage \vs\ laser pulse energy and voltage \vs\ temperature curves, estimates of the \gls{zbef} can be achieved for materials where field evaporation is thermally activated. This method represents a simple way to estimate this quantity that has been under-reported in the literature.
Using previously published data of CdTe,%
\cite{Diercks2015}
we measure an evaporation field of $F_E = \SI{18.07(87)}{\V\per\nm}$, which matches a previously published value but will smaller error.%
\cite{Diercks2018}
We also show that the assumptions permit recovery of the \gls{cp} of CdTe at \SI{23.1}{\kelvin} of $C_p = \SI{11.27(254)}{\J\per\K\per\mol}$. Comparing these values to the literature provides demonstrates that the method provides accurate estimates.
The procedure is then applied to the organic small-molecule material \gls{irppy}, yielding $F_E = \SI{7.49(96)}{\V\per\nm}$ and $C_p = \SI{173(27)}{\J\per\K\per\mol}$. This low \gls{zbef} provides a starting point for understanding how small-molecule organic semiconducting materials behave in \gls{apt}.

\begin{acknowledgments}
This work was supported by the U.S. Department of Energy, Office of Science, Basic Energy Sciences under Award DE-SC0018021 and by funding from Universal Display Corporation. The \gls{leap} was supported by a National Science Foundation Major Research Instrumentation grant DMR-1040456. Dr. David R. Diercks kindly provided the data for the CdTe analysis.%
\cite{Diercks2015}
\end{acknowledgments}

\section*{Data Availability Statement}
The data that support the findings of this study are available from the corresponding author upon reasonable request.

\glsaddall
\bibliography{biblio}
\end{document}

%% file: abbreviations.tex
\usepackage[automake,toc,nonumberlist]{glossaries}
\usepackage[symbols]{glossaries-extra}
\glssetcategoryattribute{general}{nohyper}{true}

\newglossaryentry{apt}
{
name={APT},
description={atom probe tomography},
first={atom probe tomography (\glsentrytext{apt})}
}
\newglossaryentry{leap}
{
name={LEAP},
description={Local Electrode Atom Probe\texttrademark\ 4000X Si},
first={\glsentrydesc{leap} (\glsentrytext{leap})}
}
\newglossaryentry{icf}
{
name={ICF},
description={image compression factor. A multiplicative factor that is used to approximate the deviation of particle trajectories in APT analysis from a radial projection},
first={image compression factor (\glsentrytext{icf})}
}
\newglossaryentry{oled}
{
name={OLED},
description={organic light-emitting diode},
first={\glsentrydesc{oled} (\glsentrytext{oled})},
plural={OLEDs},
firstplural={\glsentrydesc{oled}s (\glsentryplural{oled})}
}
\newglossaryentry{opv}
{
name={OPV},
description={organic photovoltaic},
first={\glsentrydesc{opv} (\glsentrytext{opv})},
plural={OPVs},
firstplural={\glsentrydesc{opv}s (\glsentryplural{opv})}
}
\newglossaryentry{otft}
{
name={OTFT},
description={organic thin-film transistor},
first={\glsentrydesc{otft} (\glsentrytext{otft})},
plural={OTFTs},
firstplural={\glsentrydesc{otft}s (\glsentryplural{otft})}
}
\newglossaryentry{mrp}
{
name={MRP},
description={mass resolving power. A characterization of the resolution of a mass spectrum in APT, it is defined as $\Delta m/m$, where $\Delta m$ is the full-width at half-maximum of the peak at mass $m$},
first={mass resolving power (\glsentrytext{mrp})},
plural={MRPs},
firstplural={mass resolving powers (\glsentryplural{mrp})}
}
\newglossaryentry{ff}
{
name={$k_f$},
sort={kf},
description={field factor. A constant which accounts for the influence of the tip shape on the electric field},
first={field factor (\glsentrytext{ff})}
}
\newglossaryentry{de}
{
name={$\eta$},
sort={eta},
description={detection efficiency. The fraction of ions detected. The efficiency is less than unity because of the microchannel plate that is used for signal amplification},
first={detection efficiency (\glsentrytext{de})}
}
\newglossaryentry{zbef}
{
name={$F_E$},
sort={zbef},
description={zero-barrier evaporation field. The field at which},
first={zero-barrier evaporation field (\glsentrytext{zbef})}
}
\newglossaryentry{fim}
{
name={FIM},
description={field ion microscopy},
first={field ion microscopy (FIM)}
}
\newglossaryentry{tem}
{
name={TEM},
description={transmission electron microscopy},
first={\glsentrydesc{tem} (\glsentrytext{tem})}
}
\newglossaryentry{fib}
{
name={FIB},
description={focused ion beam},
first={\glsentrydesc{fib} (\glsentrytext{fib})}
}
\newglossaryentry{cp}
{
name={$C_p$},
description={specific heat},
first={\glsentrydesc{cp} (\glsentrytext{cp})}
}
\newglossaryentry{irppy}
{
name={$\mathrm{Ir(ppy)_3}$},
sort={irppy},
description={tris[2-phenylpyridinato-C2,N]iridium(III)},
first={\glsentrydesc{irppy} (\glsentrytext{irppy})}
}
\newglossaryentry{ppy}
{
name={ppy},
description={2-phenylpyridine},
first={\glsentrydesc{ppy} (\glsentrytext{ppy})}
}
\newglossaryentry{alq}
{
name={$\mathrm{Alq_3}$},
sort={alq},
description={tris-(8-hydroxyquinoline)aluminum},
first={\glsentrydesc{alq} (\glsentrytext{alq})}
}
\newglossaryentry{cbp}
{
name={CBP},
description={4,4'-bis(N-carbazolyl)-1,1'-biphenyl},
first={\glsentrydesc{cbp} (\glsentrytext{cbp})}
}
\newglossaryentry{mcbp}
{
name={mCBP},
description={3,3'-Di(9H-carbazol-9-yl)-1,1'-biphenyl},
first={\glsentrydesc{cbp} (\glsentrytext{cbp})}
}
\newglossaryentry{simcp2}
{
name={SimCP2},
description={bis[3,5-di(9H-carbazol-9-yl)phenyl]diphenylsilane},
first={\glsentrydesc{simcp2} (\glsentrytext{simcp2})}
}
\newglossaryentry{dip}
{
name={DIP},
description={diindenoperylene},
first={\glsentrydesc{dip} (\glsentrytext{dip})}
}
\newglossaryentry{dcm2}
{
name={DCM2},
description={4-(dicyanomethylene)-2-methyl-6-julolidyl-9-enyl-4H-pyran},
first={\glsentrydesc{dcm2} (\glsentrytext{dcm2})}
}
\newglossaryentry{bphen}
{
name={BPhen},
description={bathophenanthroline},
first={\glsentrydesc{bphen} (\glsentrytext{bphen})}
}
\newglossaryentry{tc}
{
name={Tc},
description={tetracene},
first={\glsentrydesc{tc} (\glsentrytext{tc})}
}
\newglossaryentry{bpc}
{
name={BPC},
description={4-(N-carbazolyl)biphenyl},
first={\glsentrydesc{bpc} (\glsentrytext{bpc})}
}
\newglossaryentry{dac}
{
name={DAc},
description={Diels-Alder cycloadduct},
first={\glsentrydesc{dac} (\glsentrytext{dac})}
}
\newglossaryentry{tcp}
{
name={TCP},
description={1,3,5-tris(carbazol-9-yl)benzene},
first={\glsentrydesc{tcp} (\glsentrytext{tcp})}
}
\newglossaryentry{tcta}
{
name={TCTA},
description={4,4',4''-tris(N-carbazolyl)-triphenylamine},
first={\glsentrydesc{tcta} (\glsentrytext{tcta})}
}
\newglossaryentry{mcp}
{
name={MCP},
description={microchannel plate},
first={\glsentrydesc{mcp} (\glsentrytext{mcp})}
}
\newglossaryentry{ivas}
{
name={IVAS},
description={integrated visualization and analysis software. CAMECA's software package for reconstructing and analyzing APT data},
first={integrated visualization and analysis software (\glsentrytext{ivas})}
}
\newglossaryentry{fov}
{
name={FOV},
description={field-of-view},
first={\glsentrydesc{fov} (\glsentrytext{fov})}
}
\newglossaryentry{mde}
{
name={MDE},
description={multiple detection event. When multiple ions are detected during a single pulse for APT},
first={multiple detection event (\glsentrytext{mde})},
plural={MDEs},
firstplural={multiple detection events (\glsentryplural{mde})}
}

\makeglossaries